\documentstyle[aps,prl,twocolumn]{revtex}

\begin{document}

\bibliographystyle{prsty}

\author{I. Safi and H.J. Schulz}
\title{Interacting electrons with spin in a one--dimensional dirty wire connected
to leads
}
\address{Laboratoire de Physique des Solides, Universit\'e Paris--Sud,
91405 Orsay, France}

\maketitle
\begin{abstract}
We investigate a one--dimensional wire of interacting electrons connected to
semi--infinite leads in the absence and in the presence of a backscattering
potential. An electron incident on the clean wire is perfectly transmitted
into spatially separated spin and charge parts in the noninteracting leads,
a result we extend to any finite-range interactions. The backscattering
potential is renormalized in a non-universal way and therefore the reduction
in the conductance is more complicated than the laws derived up to now: it has
power laws as a function of temperature, wire length, and also the
distance of a barrier to the contacts.
\end{abstract}

\draft 


Recently intensive theoretical and experimental effort has been devoted to
the understanding of the transport properties of a Luttinger liquid. The
theoretical prediction of conductance renormalization by the interactions
has not been observed in quantum wires \cite{tarucha,yacoby_second}. The
physics of such tiny structures has shown that one has to take into
consideration the experimental setup. In that context, a model
\cite{ines,maslov_g} simulating the two-dimensional Fermi gas into which a
quantum wire opens by perfect one-dimensional leads was proposed; this
amounts to retain only the propagating mode. In that geometry, the
conductance is not renormalized by the interactions as commonly
accepted. The effect of a backscattering potential on the conductance of
spinless electrons was discussed in ref.\cite{ines,ines_nato,ines_proc_arcs}
taking into account contact effects which had been ignored in other
work\cite{maslov_disorder,furusaki_fil_fini}. In this letter, we consider
the nontrivial effects of the external leads, the contacts, and of spin. We
will study the transmission process of an incident spin-polarized electron
through the clean wire, demonstrating an interesting consequence of the
spin--charge separation typical of the Luttinger liquid. We then treat the
effect of weak impurities, showing that their effect on the conductance
depends crucially on contact effects, especially when isolated barriers are
considered. This may be important in many experimental cases as currently
studied wires are often quite short. We give an improved general framework to
treat weak disorder at finite temperature and in an inhomogeneous system.

An important low-energy property of an interacting one-dimensional system is
the separation of spin and charge
dynamics\cite{emery,schulz_revue,voit_revue}. The Hamiltonian is decoupled
as $H=H_\rho +H_\sigma$, where:
\begin{equation}  \label{Hnuspin}
H_\nu =\int \frac{dx}{2\pi }\left[ u_\nu K_\nu  \Pi_\nu
 ^2+\frac{u_\nu }{K_\nu }\left( \partial _x\Phi _\nu \right) ^2\right]
\end{equation}
for $\nu =\rho ,\sigma $. The boson fields are related to the charge and
spin density ($\rho$ and $\sigma$) through: ${\sqrt{2}} \partial _x\Phi
_\rho (x)/\pi =\rho$ and ${\sqrt{2}} \partial _x\Phi _\sigma /\pi=
\sigma$. $\Pi_\nu$ is the momentum density conjugate field to $\Phi_\nu$.
In the absence of interactions, $K_\nu=1,u_\nu=v_F$.  Consider now an
interacting wire delimited by $[-a,a]$ connected perfectly to noninteracting
leads. The global system thus formed is described by the Hamiltonian
(\ref{Hnuspin}) with spatially varying parameters $u_\nu (x)$, and $K_\nu
(x)$. We set $u_\nu =v_F,$ $K_\nu =1$ on the external leads, i.e. for
$|x|>a$. For simplicity, we consider the case $u(x)=u_\nu$ and $K(x)=K_\nu$
constants on the interacting wire.

An interesting issue we consider first is the transmission process of an
incident electron: let us imagine that we inject an electron of definite
spin in the left lead, and place a spin and charge detector at some location
on the right lead.  This corresponds to create a kink in both $\Phi _\rho $
and $\Phi _\sigma $ whose time evolution has now to be solved given the
initial conditions. The equations of motion for these fields are decoupled,
and require their continuity at the contacts as well as that of $u_\nu
\partial _x\Phi _\nu /K_\nu $. In particular, both the charge and spin
current, $ j_\nu =\frac{\sqrt{2}}\pi \partial _t\Phi _\nu $ are
conserved. This is because we restrict ourselves to interactions which
conserve these currents: we suppose that both umklapp processes and as the
backscattering of electrons with opposite spin are irrelevant, as is the
case for purely repulsive interactions.\cite{emery,schulz_revue,voit_revue}

The injected electron propagates freely in the left lead until it reaches
the first contact with the interacting wire. In view of the above continuity
requirements, its charge and spin are reflected with two different
coefficients $\gamma_\nu$ due to the change in interactions: $\gamma _\nu
=(1-K_\nu )/(1+K_\nu )$. The transmitted charge and spin propagate at
different velocities $u_\rho \neq u_\sigma $, thus reach the second contact
where they get partially transmitted at different times: $t_\nu =2a/u_\nu
$. Since the transmitted charge and spin propagate at the same Fermi
velocity in the right non--interacting lead, they will stay {\em spatially
separated}, at a distance $v_F\left| t_\rho -t_\sigma \right| $
(Fig.\ref{f:refspin}). The remaining charge (spin) in the wire will continue
to bounce back and forth at the contacts, letting at each time cycle $t_\rho
$ ($t_\sigma $) a partial charge (spin) come out. Thus we end up with a
series of spatially separated partial charge and spin spikes on the two non
interacting leads: they don't recombine as one might expect. The transmitted
spikes rather correspond to a complicated superposition of electron-hole
excitations. Note that in the relevant case of spin-invariant interactions,
$K_\sigma=1$, the spin part is not reflected but gets directly transmitted to
the right lead, while the charge undergoes the multiple reflection
process. At times very long compared to $t_\rho $ ($ t_\sigma $), the series
of transmitted charge (spin) spikes sum up to unity.  Thus the transmission
of an incident spin polarized flux is perfect.  We can generalize the
perfect transmission result to any profile of the interaction parameters,
and even more, to the case of {\em Coulomb interactions screened on the
measuring leads }\cite{these}. By arguments analogous to those of the
spinless case \cite{ines,landauer:70} the perfect transmission leads to a
conductance $g=2e^2/h$, independent of interactions. 

We will now show that in the presence of backscattering by impurities or at
the contacts the conductance depends on the interactions, but is generally
affected by the external leads. Consider any backscattering potential
$V(x)$. The most general representation of the backscattering Hamiltonian
$\delta H$ respecting the initial symmetries is:\cite{kane_fisher}
\begin{equation}  \label{termes}
\delta H\sim\int_{-a}^{+a} dx \sum_{m_\rho,m_\sigma}'
V(x;m_\rho,m_\sigma) e^{2im_\rho \phi
_0(x)}e^{i\sqrt{2}\left( m_\rho \Phi _\rho +m_\sigma
\Phi _\sigma \right) }
\end{equation}
The sum runs over integers $m_{\rho}$ and $m_{\sigma}$ of the same parity.
The function $\phi_0$ includes $k_F x$ as well as the effects of the
inhomogeneity of the interactions and the forward scattering\cite{these}.

For definiteness we specialize to the case of a barrier localized at a
point $x$. In order to derive the renormalization equations, we expand the
exact partition function $Z$ at finite temperature in terms of $V$. $Z$
describes a neutral gas of integer charges restricted to two circles, one
corresponding to the charge ($\nu =\rho $) and the other to the spin ($\nu
=\sigma $) degree of freedom. The radius of each circle $\nu $ is given by
$\beta $. Two charges on different circles don't interact, but have to be of
the same parity if located at the same angle. Two charges on the same circle
$\nu $ interact via $U_\nu (x,x,\tau )=\left\langle \Phi _\nu
(x,0)\left[\Phi _\nu(x,0)-\Phi _\nu (x,\tau )\right]\right\rangle $. $U_\nu
$ is not the Coulomb interaction, but an infinite series of Logarithmic
terms which are related to the transmission process we exposed earlier.

We increase the cutoff to $\tau =\tau _0e^l$, where $\tau _0$ is the bare
cutoff of the order of the inverse of the Fermi energy, and modify the
parameters in order to keep $Z$ invariant. The leading-order flow equations
 at finite temperature are:
\begin{eqnarray}  \label{renormalisationspin}
\frac{dV}{dl}&=& \left[1-\frac{1}{2}\left(m_\rho^2\frac{
dU_\rho}{dl}+m_\sigma^2\frac{dU_\sigma}{dl} \right)\right]V 
\end{eqnarray}
where for simplicity we have omitted the arguments. In particular, through
$U$, $V$ now acquires a dependence on the position of the barrier.

We can show that new interaction terms between the charges are generated by
the renormalization, but they are decaying faster than $U_\nu $ at long time
scales \cite{these}. Thus the equation (\ref{renormalisationspin}) is
sufficient, and can be integrated straightforwardly. In the absence of an
external energy scale, the unique limitation on increasing the cutoff comes
naturally from the fact that one has to put at least two charges on the same
circle of radius $\beta $, thus we cannot go further than $\tau =\beta
/2$. We write
$$V^2(x;m_\rho,m_\sigma;l)=V^2(x;m_\rho,m_\sigma)\prod_{\nu =\rho
,\sigma}E_\nu(x;m_\rho,m_\sigma;l)$$ where $V(x;m_\rho,m_\sigma;l)$ are the
renormalized parameters.  It would be cumbersome to give the behavior of
$E_\nu$ at arbitrary temperatures. We restrict ourselves to the two limits
of high or low temperature compared to $T_L$: here, $T_L$ is a common
notation for both temperature scales $1/t_\nu=u_\nu/2a$ for
$\nu=\rho,\sigma$.  In the limit $T\gg T_L$, the multiple reflections caused
by the change in interactions don't affect the correlation functions due to
the lack of thermal coherence along the wire. Only one reflection is felt
within a thermal length $L_T=v_F/T$ from the contacts. For $\nu=\rho$ or
$\sigma$, we denote by $t_{x\nu }=(a-|x|)/u_\nu$ the time it takes for a
charge or spin excitation emanating at $x$ to reach the closest contact. Up
to a slowly varying function, we get:
\begin{eqnarray}  \label{Vrenorm}
E_\nu&\sim& \left(T\tau
_0\right) ^{m_\nu ^2K_\nu -1}\left[\left( T\tau _0\right) ^2+\tanh {}^2\pi
Tt_{x\nu }\right]^{m_\nu ^2K_\nu \gamma_\nu/2}
\end{eqnarray}
At points far from the contacts compared to $L_T$, this expression coincides
with that obtained in an infinite Luttinger liquid with parameters $K_\nu$:
\begin{equation}  \label{lowKVmm}
E_\nu\sim\left(T\tau_0\right) ^{m_\nu ^2K_\nu -1}
\end{equation}
At the contacts, a similar expression holds substituting $K_\nu \rightarrow
K_{a\nu }=1-\gamma_\nu$, which is the charge/spin transmission coefficient
through one contact.  In the intermediate region $\tau _0\ll t_{x\nu }\ll
\beta $, the expression (\ref{Vrenorm}) has an additional power of $t_{x\nu
}$.

In the limit of low temperature compared to $T_L$, the multiple reflections
affect the correlation functions, thus the external leads determine the
temperature dependence. But there is a nontrivial dependence on the wire
length as well as on the distance of the barrier from the contacts:
\begin{eqnarray}  \label{integration}
E_\nu&\sim & \left( T\tau _0\right) ^{m_\nu ^2-1}\left( T_L\tau
_0\right)^{-m_\nu^2 \gamma_\nu}\left[1+\left(t_{\nu x}/\tau _0\right)
^2\right]^{m_\nu ^2K_\nu\gamma_\nu/2}
\end{eqnarray}
For a barrier in the bulk, so that $t_{x\nu }\sim t_\nu/2=a/u_\nu $, this
expression simplifies to:
\begin{equation}  \label{lowVmmspin}
E_\nu\sim \left( T\tau_0\right)^{m_\nu ^2-1}
\left(T_L\tau_0\right)^{m_\nu ^2(K_\nu -1)}
\end{equation}
At the contacts, one has to replace $K_\nu $ by $K_{a\nu }$. For $t_{\nu
x}\gg \tau _0$, the expression (\ref{integration}) contains a power law in
$t_{\nu x}.$

We now discuss the effect of the backscattering potential on the
conductance. For this purpose, we use the equation of motion of $\Phi _\rho
$ to derive a Dyson equation for the nonlocal conductivity and thus cast the
exact expression for the conductance into a form appropriate for a
perturbative computation \cite{ines_nato,these}. The latter is equivalent to
inserting the renormalized parameters $V^2(x;m_\rho,m_\sigma;l)$ in the
expression for the conductance because the interactions are not
renormalized. The dominant correction is reported in table
\ref{f:spinbarriere}. It is given by $V^2(x,1,1)$, i.e. by the $2k_F$
backscattering, as long as $K_\rho>1/3$. Consider now the case
$K_\rho<1/3$. In an infinite Luttinger liquid the tendency towards a $4k_F$
CDW dominates over the $2k_F$ CDW, which is reminiscent of a tendency
towards a Wigner crystal. Consequently, in this parameter region the $4k_F$
scattering process ($m_\rho=2$, $m_\sigma=0$), corresponding to a coherent
backscattering of two electrons, becomes important.  It turns out that the
measuring leads affect this tendency. The dominant correction to the
conductance is not necessarily given by the $4k_F$ backscattering term as in
an infinite Luttinger liquid. In particular, below $T_L$ the noninteracting
leads dominate, saturating the $4k_F$ scattering. Details of the crossover
between dominant $2k_F$ and $4k_F$ scattering depend on the temperature as
well as on the location, as shown in table \ref{f:spinbarriere}. In
particular, the crossover temperature is position--dependent. For
$1/5<K_\rho<1/3$, this crossover takes place on the bulk but not near the
contacts. Thus, the external leads {\em suppress the importance of $4k_F$
scattering at enough low temperature or close to the contacts} and also
saturate the tendency towards the formation of a Wigner crystal.

All the previous results concerning one barrier, including the
renormalization equations (\ref{renormalisationspin}), can be extended to
any {\em potential $V$ with a spatial extension much less than the thermal
length $L_T$ and the wire length $L$} \cite{these}. Note that in the
presence of a potential with any extension, the partition function describes
a gas of integer charges confined to two cylinders. The renormalization
equations (\ref{renormalisationspin}) are still valid. Nevertheless, the
potential $V$ can renormalize the interactions. The computation of the
conductance can be accomplished in a perturbative way \cite{these} as long
as the potential is either weak enough or not too much extended to not
renormalize the interactions.  Now we perform the average over random
impurities extended all over the wire. For simplicity, we limit ourselves to
the case of a Gaussian distribution: $ \left\langle V(x)V(y)\right\rangle
=v_F^2\delta (x-y)/l_e$.  We perform the average of the conductance
expression. Upon integration over the wire, the reduction $\delta g$
contains an infinity of powers in the temperature. We skip the details, and
give the leading correction in table \ref{t:spindesordre}. There appear
essentially two terms in $\delta g$: the contribution of the impurities
close to the contacts, governed by the local parameter $K_{a\rho }$, and
that of the impurities on the bulk governed by $K_\rho $. The latter
dominates the former at any temperature if $K_\rho <1+\sqrt{2}$ or at
$T>T_b$ if $K_\rho >1+\sqrt{2}$, with
\begin{equation}  \label{eq:Tbulk}
T_b \tau_0=\left(T_L\tau_0\right) ^{-1/(\gamma_\rho K_\rho)}.
\end{equation}
 For $K_\rho <1/3$, the $4k_F$ backscattering term dominates only at high
temperature $T>T_L$.

Let us discuss the conductance fluctuations.  We can evaluate the variance
of the conductance, and show that:
\begin{equation}  \label{variance}
\frac 12(\delta g)^2 \leq 
\langle g^2 \rangle - \langle g \rangle^2 \leq (\delta g)^2
\end{equation}
As long as $\delta g$ is small, the conductance is self-averaging, but its
fluctuations are of the same order as its reduction.  

Let us compare finally the conductance we get to the usual conductance of an
infinite dirty Luttinger liquid. While the ballistic conductance we found is
not renormalized by the interactions, its reduction in the presence of
impurities is similar concerning its power law behavior controlled by the
interactions, but is more complicated than this in
ref.\cite{apel_rice,kane_fisher,ogata}. In particular, we must comment on
the commonly used scaling argument used to infer the properties of the
finite wire from those computed in an infinite wire: The energy cutoff
introduced by the temperature $T$ is replaced by the higher one introduced
by the wire length $T_L$. In our model, we observe that the substitution
$T\rightarrow T_L$ does not allow to go from (\ref{Vrenorm}) to
(\ref{integration}) unless $m_\rho=m_\sigma=1$ and $t_{x\nu}\sim
t_\nu$. Thus even in the presence of a barrier at a distance from the
contact of the order of the wire length, we recover the correction to the
conductance of an infinite wire (extrapolated on the whole range of
temperature) only for $K_\rho>1/3$ (see table \ref{f:spinbarriere}). In the
presence of extended disorder, table \ref{t:spindesordre} shows that the
infinite wire result is recovered merely in the window $1/3<K_\rho
<1+\sqrt{2}$.

We come now to the comparison with experiment. Recently, Tarucha {\em et
al.} studied relatively clean and long (up to $10\mu $m, corresponding to
$T_L = 0.1$K) quantum wires fabricated in an AlGaAs/GaAs heterostructure. At
high enough temperature, the conductance is quantized in units of
$2e^2/h$. As the temperature is lowered, the impurities become more
efficient, and the quantization less perfect. The reduction of the
one-channel wire conductance with the temperature is attributed to
electron-electron interactions and fitted with a power law controlled by a
parameter $K_\rho =0.7$. These experimental observations contradict the
result of renormalization of the ballistic conductance by the
interactions. Instead, they can be explained in a natural way by the model
we proposed. Let's discuss further the backscattering potential. It can
emanate either from impurities or from imperfections at the opening of the
wire in the two-dimensional gas. Our model yields a peculiar dependence of
the reduction to the conductance on the temperature and the location of the
barriers. For instance, if the backscattering is due merely to the contacts,
the parameter Tarucha {\em et al.} infer would correspond to the local
parameter $K_{a\rho }=2K_\rho/(1+K_\rho)$, which yields $K_\rho =0.5$ . On
the other hand, they don't have a good agreement with the intuitive law
$[T^2+T_L^2]^{(K_\rho -1)/2}$ \cite{ogata}. We predict instead different
possible functions of the temperature. For instance, in the presence of
impurities in the intermediate region, we get two successive power laws
controlled first by $K_\rho $ then by $K_{a\rho }$ upon cooling (table
\ref{f:spinbarriere}). We are more inclined to discard the possibility of a
Gaussian distribution extended over the wires studied by Tarucha and
al. First, the fluctuations in the conductance are not of the same order of
magnitude as its reduction, eq.(\ref{variance}). Secondly, the reduction
depends on the gate voltage, and shows even some oscillations for the $10\mu
m$ wire. Probably, the backscattering potential configuration depends on
each wire, and more precise fit and evaluation of $K_\rho $ are required. In
particular, one hopes to infer a parameter $K_\rho$ depending on the filling
factor.

Another interesting experiment was performed more recently by Yacoby {\em et
al.}  \cite{yacoby_second}.  The conductance is again not
renormalized by the interactions at high enough temperature. Upon cooling,
the conductance decreases, the reduction being more important in the longer
wires. This is qualitatively coherent with the predictions we get in the
presence of backscattering for $K_\rho <1$. Nonetheless, the observed
reduction in the conductance is independent on the gate voltage, while both
$K_\rho$ and the overall scale of $\delta g$ (which depends on the Fermi
energy) are expected to depend on the gate voltage.  As discussed in
Ref.\cite{yacoby_second}, it's not granted that the leads have a non-Fermi
behavior in such sophisticated experimental set up, and a more suited model
is needed.

To conclude, we studied the properties of an interacting wire connected to
measuring leads in the absence and in the presence of a back scattering
potential. The clean wire separates an incident electron into spin and
charge parts, and this separation persists even in the noninteracting leads.
The total transmission is perfect for any range of the interactions. In the
presence of a backscattering potential, the conductance is affected both by
the interactions and by the external leads. It's only for certain impurity
distributions and interaction parameters that the latter don't intervene.

\begin{figure}[htb]
\caption{Dynamic transmission of an incident electron with spin up. The
charge and spin are separated even in the noninteracting leads. As an
example, we consider here $u_\rho>u_\sigma$ 
and $K_\sigma<1$, $K_\rho>1$.}
\label{f:refspin}
\end{figure}

\begin{figure}[htb]
\caption{The reduction to the conductance scaled by the backscattering
strength as a function of the temperature for different potential
distributions. We choose $K_\rho=0.5$, $K_\sigma=1$, and $E_F=148 T_L$, thus
$\ln (T_L/E_F)=5$. The dotted, dashed and dot--dashed curves correspond
respectively to a barrier at the center of the wire, at the contact and at
an intermediate point $x$. There is a crossover at $T_L$ from a power law
controlled by $K_\rho$ ($K_{a\rho}$) at the center (at the contact) to a
plateaus. At $x$, there are two crossovers: one at $T_x=u_\rho/t_x$ from
$T^{K_\rho-1}$ to $T^{K_{a\rho}-1}$, than a saturation occurs at $T_L$. The
continuous curve correspond to an extended Gaussian distribution, where the
power law at $T<T_L$ is governed by $K_\rho$ since the interactions are
repulsive. Since $K_\rho=0.5>1/3$, the $2k_F$ backscattering dominates at
all $T$. }
\label{f:allcourbes}
\end{figure}

\begin{table}[hbt]
\caption{A weak barrier at $x$: dominant reduction of the conductance
depending on the temperature and on $K_\rho$ for the case
$K_\sigma=1$. $T_x=\pi/t_x$ is the inverse of the time to get from $x$ to
the closest contact. Every temperature has to be multiplied by $\tau_0$. The
first line gives the behavior all over the wire at $T<T_L$, governed only by
the $2k_F$ backscattering. Note that the second line has to be skipped in
case $T_x\sim T_L$. The last line concerns only $t_x\gg \tau_0$.}
\begin{tabular}{cccc}
&$K_\rho<1/5$&$1/5<K_\rho<1/3$&$1/3<K_\rho$\\ \tableline $T<T_L$
&\multicolumn{3}{c} {$T_L^{-\gamma_\rho}T_x^{-\gamma_\rho K_\rho}$}\\
$T_L<T< T_x$
&$T^{2-4\gamma_\rho}T_x^{-4K_\rho\gamma_\rho}$&\multicolumn{2}{c}
{$T^{-\gamma_\rho}T_x^{-\gamma_\rho K_\rho}$}\\ $T_x<
T$&\multicolumn{2}{c}{$T^{4K_\rho-2}$}&$T^{K_\rho-1}$\\
\end{tabular}
\label{f:spinbarriere}
\end{table}

\begin{table}[hbt]
\caption{Gaussian disorder: dominant reduction of the conductance $\delta g$
multiplied by $l_e$. The $4k_F$ backscattering dominates for $K_\rho<1/3$
only at high temperature $T>T_L$. The temperature $T_b$ is given by
eq. (\ref{eq:Tbulk}).}
\begin{tabular}{cccc}
&$K_\rho<1/3$&$1/3<K_\rho<1+\sqrt{2}$&$1+\sqrt{2}<K_\rho$\\
\tableline
$T<T_L$ &\multicolumn{2}{c} {$T_L^{K_\rho-2}$}&$T_L^{-\gamma_\rho}$\\
$T_L<T<T_b$ &$T^{4K_\rho-2}/T_L$& $T^{K_\rho-1}/T_L$&$ T^{-\gamma_\rho}$\\
$T_b<T$&$T^{4K_\rho-2}/T_L$&\multicolumn{2}{c}{$T^{K_\rho-1}/T_L$}\\
\end{tabular}
\label{t:spindesordre}
\end{table}

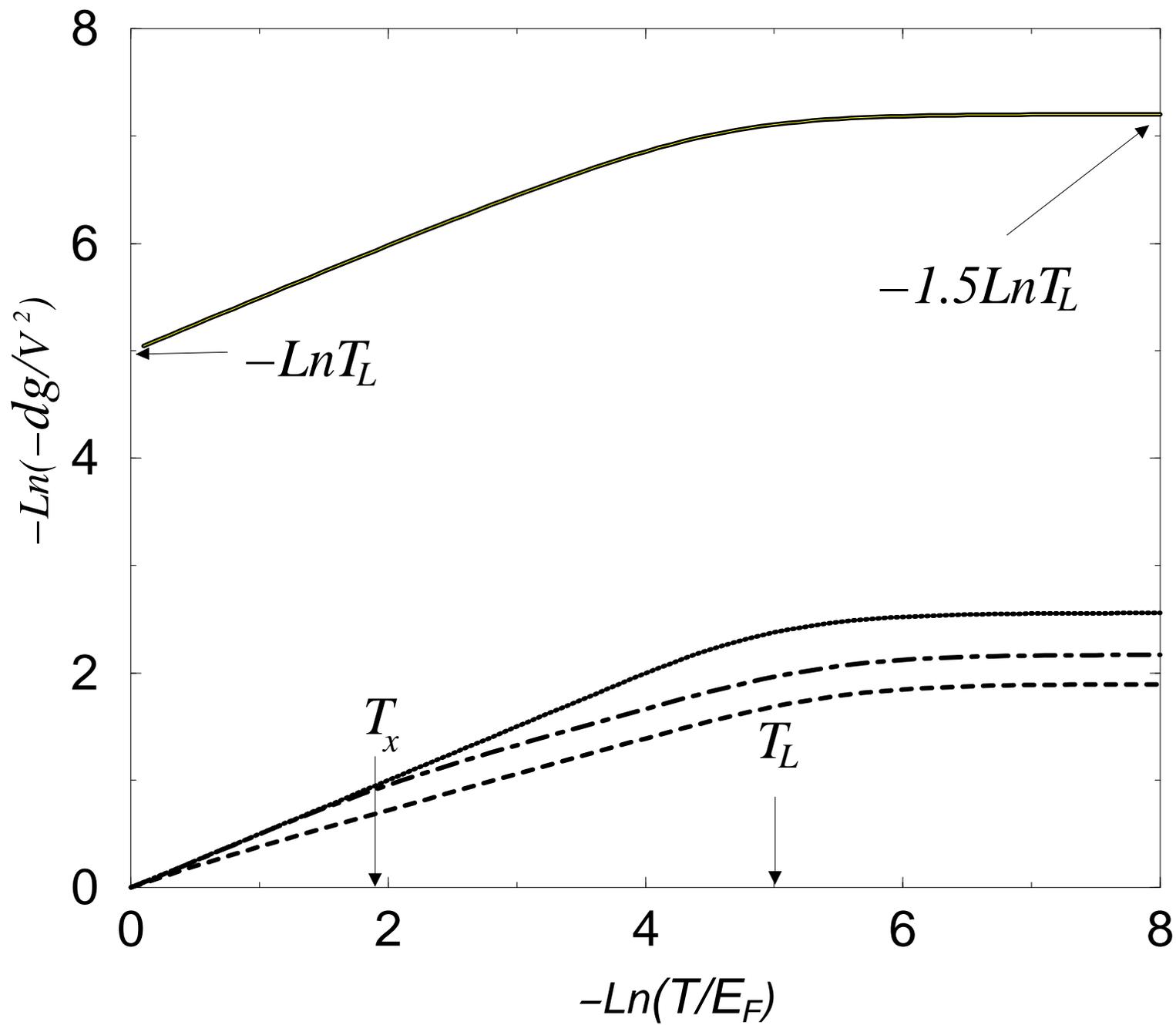
\end{document}